\documentclass[showpacs, pra,onecolumn,preprintnumbers ,amsmath, amssymb, superscriptaddress, aps]{revtex4-2}
\usepackage{color}
\usepackage{amsmath,amssymb}
\usepackage{pifont}
\usepackage{amssymb}  
\usepackage{bbold}
\usepackage{float}
\usepackage{subfloat}
\usepackage[squaren]{SIunits}

\usepackage[caption=false]{subfig}
\usepackage{tikz}
\usepackage{makecell}
\usepackage{subfig}
\usepackage{pifont}   
\usepackage{graphicx} 
\graphicspath{{Figures/}}
\usepackage{dcolumn}  
\usepackage{bm}       
\usepackage{multirow} 
\usepackage{placeins}
\usepackage[colorlinks]{hyperref}
\usepackage{mathtools}
\usepackage{appendix}

\newcommand{\rr} {\color{red}}

\begin{document}
	
	\title{Transmission gaps in phosphorene superlattice}
	\date{\today}
	
	\author{Jilali Seffadi}
		\affiliation{Laboratory of Theoretical Physics, Faculty of Sciences, Choua\"ib Doukkali University, PO Box 20, 24000 El Jadida, Morocco}
		\author{Ilham Redouani}
		\affiliation{Laboratory of Theoretical Physics, Faculty of Sciences, Choua\"ib Doukkali University, PO Box 20, 24000 El Jadida, Morocco}
		\author{Youness Zahidi}
		\affiliation{MRI Labortory, Polydisciplinary Faculty, Sultan Moulay Selimane University,
			Khouribga, Morocco}
	\author{Ahmed Jellal}
	\email{a.jellal@ucd.ac.ma}
	\affiliation{Laboratory of Theoretical Physics, Faculty of Sciences, Choua\"ib Doukkali University, PO Box 20, 24000 El Jadida, Morocco}
	\affiliation{Canadian Quantum  Research Center,
		204-3002 32 Ave Vernon,  BC V1T 2L7,  Canada}

	
	\begin{abstract}
Our research focuses on the transmission gaps of charge carriers passing through phosphorene superlattice, which are made up of a series of barriers and wells generating $n$ identical cells. We determine the solutions of the energy spectrum  and then transmission using  Bloch's theorem and the transfer-matrix approach. The analysis will be done on the impact of incident energy, barrier height, potential widths, period number, and transverse wave vector on transmission. We show that 	pseudo-gaps appear and turn into real transmission gaps by increasing the number of cells. Their number, width, and position can be tuned by changing the physical parameters of the structure. At normal incidence, a forbidden gap is found, meaning that there is no Klein tunneling effect, in contrast to the case of graphene. Our findings can be used to create a variety of phosphorene-based electronic devices. 
		
	\end{abstract}

	\maketitle

	\section{Introduction}
	
There has been an increase in interest in two-dimensional (2D) layered materials \cite{ref3} since the discovery of graphene, a single layer of carbon atoms organized in 2D honeycomb lattice \cite{ref1}. Several other 2D materials, including transition metal dichalcogenides, silicene \cite{ref1-}, germanene \cite{ref2-} and borophene \cite{ref3-3}, have created sensations in the world of nanotechnology. These 2D materials show a number of unique physical properties that are not present in their bulk counterparts, which make them promising candidates for electronic applications \cite{Novoselov05,Lee10}.  Graphene, as the most well-known 2D material, has no electronic band gap in its electronic structure. However, the lack of a gap has been considered an impediment in practical applications of semiconductor device design \cite{Castro09}. As a result, it is natural to look for other 2D crystalline materials beyond graphene with a gap in their band structures.

One of the most promising 2D materials is phosphorene, which is a black phosphorus allotrope of phosphorus \cite{Li14, Rodin14, Koenig14, Liu14}. It is a layered material in which weak Van der Waals interactions stack individual atomic layers on top of one another. The phosphorus atom forms a puckered hexagonal lattice in each layer by covalently bonding with three nearby atoms via the $sp^3$ hybridization process.  In 2014, it was successfully separated from its layered bulk using the mechanical exfoliation technique \cite{Li14}. This report has induced heightened research interest in its physical properties and possible applications. According to \cite{Qiao14}, this novel 2D material has an adjustable band gap that varies from approximately 0.5 eV for five-layer structures to 1.5 eV for monolayer ones. Phosphorrene has been intensively investigated due to its special, extremely anisotropic electronic and optical characteristics.  Among phosphorus allotropes, it is the most stable and exhibits strong reactivity in ambient circumstances \cite{Nishii87, Castellanos14}.

Theories and experiments have been used to examine different aspects of phosphorene, e.g. transport properties \cite{Liu14,ref10,Seffadi22}, strain modification \cite{ref11,ref12}, field transistor \cite{Li14,ref14,ref15}, optoelectronics and electronics \cite{ref16,ref17,ref18,ref19,ref20}, heterostructures and PN junctions \cite{ref21,ref22,ref23}, and excitons \cite{ref24,ref25}. They have confirmed the theory and observed a strong in-plane anisotropy \cite{Qiao14}, anisotropic Rashba spin-orbit coupling \cite{Popovi15}, anisotropic Landau levels \cite{ref20}, a changeable bandgap \cite{Qiao14, Li14} and high mobility \cite{Liu14}. The creation of electronic and optical devices can take advantage of these intriguing features. A superlattice design concept has drawn growing attention from researchers as a means of creating these high-performance applications based on their distinctive and unusual features.  
For instance, external periodic electric and/or magnetic fields can be used to create superlattice structures.  These structures display a variety of fascinating phenomena, such as negative differential conductivity and gap openings near the mini-Brillouin zone boundary. 
Superlattice structures for graphene are made using a variety of techniques, as reported, for instance, in \cite{Uddin14, zahidi16, Maksimova12, S.Uddin15, Jellal15}.

	
We examine the transmission gap of charge carriers passing through a monolayer phosphorene superlattice with periodic potential patterns in part because of the findings in graphene superlattices \cite{ref55, ref56, Cervantes16}, which served as our inspiration. We first determine the eigenvalues and eigenspinors by solving the continuum model's eigenvalue equation for one unit cell. Subsequently, for superlattice, we derive the electronic dispersion  and the transmission as a function of the physical parameters using Bloch's theorem and the transfer matrix approach. Our numerical findings demonstrate that the periodic potential structure has several transmission gaps. Furthermore, as the number of cells increases, certain pseudo-gaps that were previously transmission become actual transmission gaps. Discussion is also had over how the physical characteristics of energy, barrier width, transversal wave vector, barrier height, and number of  cells affect transmission gaps. In addition, we discover that, unlike gated graphene superlattices, there is no Klein tunneling at normal incidence because there is a forbidden gap.


The structure of this work is as follows: in Section  \text{\ref{Sec2}}, we define the phosphorene superlattice by choosing the proper Hamiltonian, which will be utilized to determine the solutions of the energy spectrum for one unit cell. Then, using the Bloch's theorem and transfer matrix approach, we determine the dispersion relation and transmission for the superlattice. We numerically analyze the transmission  under various conditions of the physical parameters  in Section \text{\ref{Sec3}}. Finally, we close by summarizing and  concluding our results.

	\section{Theoretical model}\label{Sec2}

\begin{figure}[h!]
		\centering
		\centering{\includegraphics[scale=0.5]{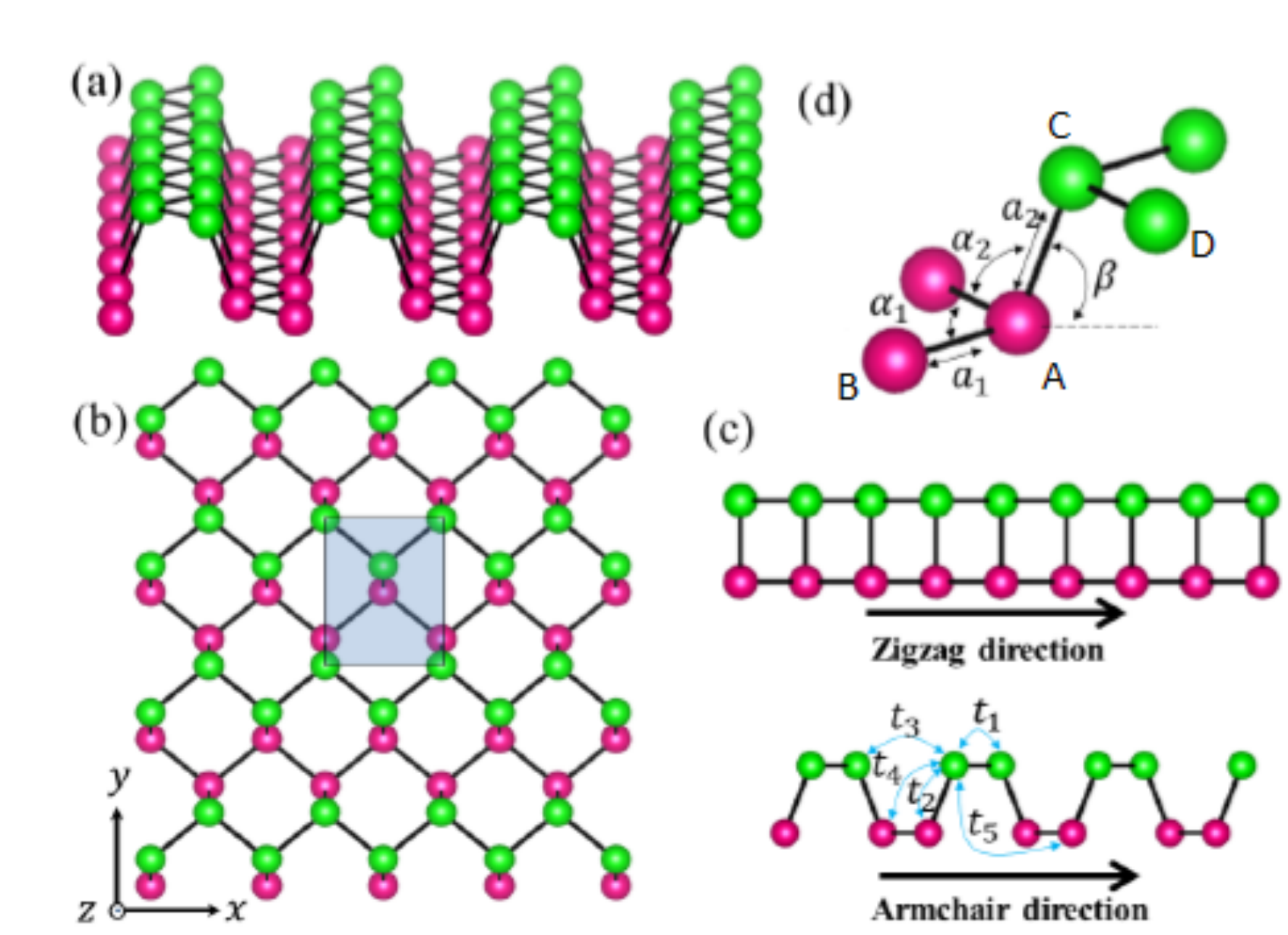}}
	\caption{(color online) {(a): 
	A monolayer phosphorene lattice structure, with each monolayer representing the phosphorus atoms at various sublayers with a different color.
	(b): 2D monolayer lattice that has been projected onto the $ xy $-plane.
	The unit cell is shown by a dashed rectangle.
	(c): Interatomic coupling at the zigzag and armchair edges.
	(d): Monolayer phosphorene  lattice structure and tightly-binding model parameters. }}\label{f05} 			
\end{figure}

In contrast to graphene, as illustrated in  Fig. \ref{f05}{\rr a} phosphorene is a single atomic layer of black phosphorus that, due to its $sp^3$ hybridization, creates puckered honeycomb-structured layers \cite{Rodin14, Castellanos14}. These two atomic layers are held together by weak van der Waals forces \cite{Appalakondaiah12}. The unit cell of phosphorene has a puckered structure and four phosphorus atoms. 
 Each sublayer contains two atomic layers, as depicted in Fig. \text{\ref{f05}}{\rr b}. The distances of the top and bottom atoms ($a _2=2.24\ \AA$) and the two nearest atoms ($a_1=2.22\ \AA$) are marginally different.

\begin{figure}[ht]
	\centering
	\includegraphics[width=14.4cm, height=4.6cm]{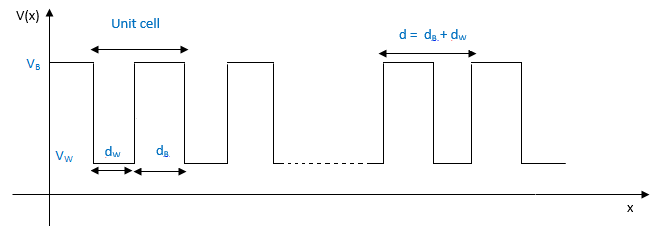}
	\caption{(color online) Periodic potentials are applied to phosphorene to generate a superlattice. }\label{f01}
\end{figure}

In the context of the low-energy 5-hopping parameter tight-binding method, the Hamiltonian for monolayer phosphorene can be stated as follows in momentum space \cite{Popovi15, Pereira15, Sarkar17, Ezawa14}: 
\begin{equation}
	H(k)=\left(%
	\begin{array}{cc}
		u_{0}+ \eta_{x} k_{x}^{2}+\eta_{y} k_{y}^{2} & \delta + \gamma_{x} k_{x}^{2}+\gamma_{y} k_{y}^{2}+i\chi k_x \\
		\delta+ \gamma_{x} k_{x}^{2}+\gamma_{y} k_{y}^{2}-i\chi k_x & u_{0}+ \eta_{x} k_{x}^{2}+\eta_{y} k_{y}^{2}\\
	\end{array}%
	\right)
\end{equation}
where the parameters $u_{0}=\delta_{AA}+ \delta_{AD}$, $\delta=\delta_{AB}+ \delta_{AC}$, $\eta_{x}=\eta_{AA}+ \eta_{AD}$, $\eta_{y}=\gamma_{AA}+ \gamma_{AD}$, $\gamma_{x}=\eta_{AB}+\eta_{AC}$, $\gamma_{y}=\gamma_{AB}+\gamma_{AC}$ and
$\chi=\chi_{AB}+ \chi_{AC}$ have been defined. 
Table \ref{bc} provides an overview of the coefficient values of the expanded structural factors for both five-hopping models. As shown in Fig.\ref{f01}, we analyze the transmission of charge carriers in phosphorene by taking into account a periodic one-dimensional potential along the $ x $-direction. We point out that this structure can be created by supplying the phosphorene layer with a local top gate voltage \cite{Huard07}. The potential barrier height $V_B$ with width $d_B$ and potential well height $V_W$ with width $d_W$ define the one-dimensional superlattice potential structure. The width of the alternating barriers and wells is defined as $d= d_B + d_W$. The potential profile has the following form:
\begin{equation}
	\label{eq 2}
	V_j(x)=\left\{\begin{array}{llll}
		{V_B} & \mbox{if} &{id<x<id+d_B}\\
		{V_W} & \mbox{if} &  {id+d_B<x<(i+1)d}
	\end{array}\right.
\end{equation} 
where $j=B, W$ and $ i \in \mathbb{N}$. 
\begin{table}[htbp]
	\center
	\begin{tabular}{ccccc}
		\hline
		& 5-hopping    &  &5-hopping& Units \\
		\hline
		$\delta_{AA} $& 0.00& $\delta_{AB} $ &-2.85& eV \\
		$\delta_{AC} $&  3.61   &$\delta_{AD} $&-0.42 & eV \\
		$\eta_{AA}   $&  0.00   &$\eta_{AB}   $&3.91  & eV ${\AA}^{2}$ \\
		$\eta_{AC}   $&-0.53 & $\eta_{AD}   $&0.58 & eV ${\AA}^{2}$ \\
		$\gamma_{AA}$ & 0.00  & $\gamma_{AB}$ & 4.41 & eV ${\AA}^{2}$ \\
		$\gamma_{AC}$ &  0.00  &$\gamma_{AD}$&1.01  & eV ${\AA}^{2}$ \\
		$\chi_{AB} $  &  2.41  & 	$\chi_{AC}$ & 2.84& eV ${\AA}$ \\
	\end{tabular}
	\caption{Values of structure factor coefficients.}\label{bc}
\end{table}

We take into account that a phosphorene superlattice comprises of $n$ identical cells that are sandwiched between the incident and transmitted zones. Each region $j$ of the unit cell can be described by the following Hamiltonian
\begin{equation}
	H_{j}(k)=H(k)+V_{j}(x)\mathbb{I}_2.
\end{equation}
In all regions, the eigenspinors of these quasi-particles can be expressed as $\psi_j(x, y) =\phi_j(x)e^{ik_yy}$ due to momentum conservation in the $y$-direction. After that, the energy can be determined by computing the dispersion relations in region $j$
\begin{align}
	E_{j}=V_j+u_{0}+\eta_{x}k_{jx}^{2}+\eta_{y}k_{y}^{2}+s_{j}\sqrt{\left(\delta+ \gamma_{x} k_{jx}^{2}+\gamma_{y} k_{y}^{2}\right)^2+\left(\chi k_{jx}\right)^2}\label{wave10}
\end{align}
with $s_{j}={\mbox{sign}}{\left(E_j-V_j-u_{0}-\eta_{x}k_{jx}^{2}-\eta_{y}k_{y}^{2}\right)}$. It is worth noting that the part of the energy spectrum proportional to $ k_{jx}^{2}<1$ can be ignored due to the presence of $u_{0}, \delta, k_{jx}\gg k_{jx}^{2}$ at low moments. According to \eqref{wave10}, the longitudinal wave vectors $k_{jx}$ are provided by 
\begin{align}
	k_{jx}=\chi^{-1}\sqrt{\left(E_j-V_j-u_{0}-\eta_{y}k_{y}^{2}\right)^2+\left(\delta+\gamma_{y} k_{y}^{2}\right)^2}\label{wavevector12}.
\end{align}
When $V=0$, the energy bands are as follows for the incident and transmitted regions: 
\begin{align}
	E_{0}=u_{0}+\eta_{x}k_{0x}^{2}+\eta_{y}k_{y}^{2}+s_{0}\sqrt{\left(\delta+ \gamma_{x} k_{0x}^{2}+\gamma_{y} k_{y}^{2}\right)^2+\left(\chi k_{0x}\right)^2}\label{wavevector10}
\end{align}
associated with the wave vector
\begin{align}
	k_{0x}=\chi^{-1}\sqrt{\left(E_0-u_{0}-\eta_{y}k_{y}^{2}\right)^2+\left(\delta+\gamma_{y} k_{y}^{2}\right)^2}\label{wavevector12}.
\end{align}
and we have set $s_{0}={\mbox{sign}}{\left(E_0-u_{0}-\eta_{x}k_{0x}^{2}-\eta_{y}k_{y}^{2}\right)}$.
To find the corresponding eigenspinors, solve the eigenvalue equation
$H_j\phi_j=E_j\phi_j$. The outcomes of this approach are the solutions in each region
\begin{align}
	\label{eqM}
	\phi_{\sf j}&=G_j\cdot A_j
\end{align}
in which the matrices are provided by 
\begin{align}
	G_j=
	\begin{pmatrix}
		e^{ik_{j}x}& e^{-ik_{j}x}\\
		z_{j}e^{ik_{j}x}& z_{j}^{-1}e^{-ik_{j}x}
	\end{pmatrix},  \qquad
	A_{j}=
	\begin{pmatrix}
		{a_{j}} \\
		{b_{j}} \\
	\end{pmatrix}
\end{align}
including the complex number and angle 
\begin{align}
z_{j}=s_{j}e^{-i\theta_{j}}, \qquad \theta_{j}=\arctan\left( \frac{\chi k_{jx}}{\delta+\gamma_{y} k_{y}^{2}}\right)	
\end{align}
where  $s_{j}={\mbox{sign}}{\left(E_j-V_j-u_{0}-\eta_{y}k_{y}^{2}\right)}$, $a_{j}$ and $b_{j}$ are constant coefficients. Keep in mind that the solutions above are for a single primitive cell.

We apply Bloch's theorem to the superlattice and, after a simple calculation \cite{Arovas10,Barbier10,Lima15}, we derive the following  dispersion relation for infinite periodic structures 
\begin{equation}\label{f0004}
	\begin{split}
		\cos(k_xd)=\cos(d_B k_{Bx})\cos(d_W k_{Wx})+G\sin(d_B k_{Bx})\sin(d_W k_{Wx})
	\end{split}
\end{equation}
where we have set 
\begin{equation}\label{f0005}
	\begin{split}
		G=\frac{s_B s_W\cos{\theta_B}\cos{\theta_W}-1}{s_B s_W\sin{\theta_B}\sin{\theta_W}}
	\end{split}
\end{equation}
including the unit cell's length $d$ and the Bloch wave vector $k _x$.  \eqref{f0004} really displays many features that will be utilized to examine the transmission of the superlattice. 

We will then determine the  transmission that electrons will successfully pass through the phosphorene superlattice. We are particularly interested in the normalizing coefficients, components of $A_j$, on either side of the superlattices while performing this. For example, we have the incident and transmitted regions
\begin{align}
A_{in}=
\begin{pmatrix}
	{1} \\
	{r} 
\end{pmatrix}, \qquad A_{tr}=
\begin{pmatrix}
	{t} \\
	{0} 
\end{pmatrix}
\end{align}
where the coefficients for transmission and reflection, respectively, are $t$ and $r$. Imposing the continuity of the wave functions yields the coefficients $r$ and $t$. The wave functions at the interfaces of several regions should match for this to be realized. The transfer matrix formalism, as used in  \cite{You95}, is the most practical way to express this process. As a result, the continuity of the wave functions allows us to up with
\begin{align}
\begin{pmatrix}
	{1} \\
	{r} 
\end{pmatrix}
=D\cdot \begin{pmatrix}
	{t} \\
	{0} 
\end{pmatrix}
\end{align}
where $D$, the transfer matrix, is provided by 
\begin{equation}\label{eqTM}
D=G_{in}^{-1}[0]\cdot L^{n} \cdot  G_{tr}[nd]= 
\begin{pmatrix}
	D_{11} & 	D_{12} \\
	D_{21}& 	D_{22}
\end{pmatrix}
\end{equation}
in which $L$ reads as
\begin{align}
L=
G_B[0]\cdot G_{B}^{-1}[d_{B}]\cdot G_W[d_{B}]\cdot G_{W}^{-1}[d]
\end{align}
After some  algebra, we demonstrate that the transmission probability may be expressed as 
\begin{align}
T=|t|^2=\frac{1}{|D_{11}|^2}
\end{align}
To glean further information from these results and in accordance with the core components of the current system, numerical analysis will be conducted.

\section{Results and discussions }\label{Sec3}

\begin{figure}[ht]
	\centering
	\subfloat[$n=1$]{
		\centering
		\includegraphics[scale=0.5]{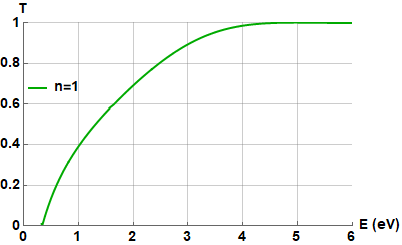}\label{f3.a}
	}\hspace{1cm}
\subfloat[$n=3$]{
		\centering
		\includegraphics[scale=0.5]{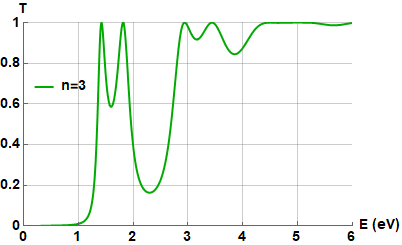}\label{f3.b}
	}\\
\subfloat[$n=5$]{
		\centering
		\includegraphics[scale=0.5]{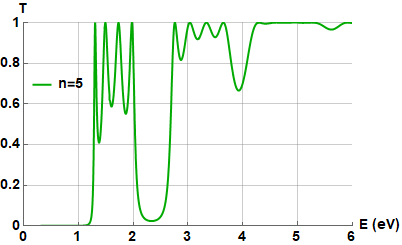}\label{f3.c}
	}\hspace{1cm}
\subfloat[$n=10$]{
		\centering
		\includegraphics[scale=0.5]{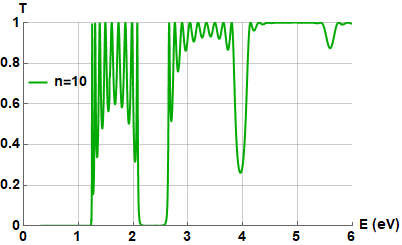}\label{f3.d}
	}\\
	\subfloat[$n=20$]{
		\centering
		\includegraphics[scale=0.5]{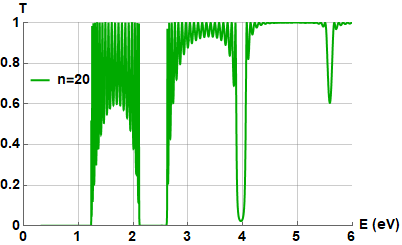}\label{f3.e}
	}\hspace{1cm}
\subfloat[$n=100$]{
		\centering
		\includegraphics[scale=0.5]{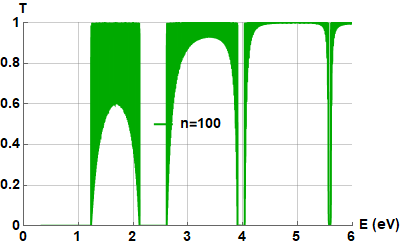}\label{f3.f}
	}
	\caption{The transmission  $T$ versus of the incident energy $E $ for $k_y=0$, $d_B=d_W=5 \ \AA$, $V_B=2$ eV and $V_W=0$ eV. 
		\textbf{\color{red}{(a)}}: $n=1$, \textbf{\color{red}{(b)}}: $n=3$, \textbf{\color{red}{(c)}}: $n=5$, \textbf{\color{red}{(d)}}: $n=10$, \textbf{\color{red}{(e)}}: $n=20$, \textbf{\color{red}{(f)}}: $n=100$. 
	}\label{f3}
\end{figure}

Fig. \text{\ref{f3}} depicts the transmission  $T$ versus the incident energy $E$ for various numbers of cells $n=1, 3, 5, 10, 20, 100$. 
This structure collapses into a single barrier with no transmission gap when $n=1 $, as shown in Fig.  \text{\ref{f3.a}}, which is due to the choice of a narrow barrier width ($d_B=5\ \AA$). In fact, the evanescent wave has a slight chance of passing through the barrier structure for small values of $d_B$. This result is similar to that found for a single barrier structure in monolayer graphene \cite{ref55}, while it is different from the results for a single barrier in monolayer graphene \cite{Chen09}. We notice that some pseudo-gaps emerge for both n=3 and n=5, respectively. For $n=100$, Fig. \text{\ref{f3.f}} clearly illustrates how these pseudo-gaps deepen as the number of cells increases, their edges steepen, and some of them develop into true transmission gaps. This behavior is comparable to that of gated graphene superlattices \cite{ref55}  and massive-massless graphene superlattices \cite{ref56}. Additionally, we observe that as $n$ grows, more oscillations outside of transmission gaps appear until total transmission is reached.  We see that the number of resonances is inversely correlated with the number of cells, which is consistent with the findings in \cite{ref55,ref56}. Furthermore, the width and location of the transmission gaps are not affected by the number of  cells, but it   has an effect on the pseudo-gaps.

\begin{figure}[ht]
 	\centering
 	\subfloat[$k_y=0\ \AA^{-1}$]{
 		\centering
 		\includegraphics[scale=0.5]{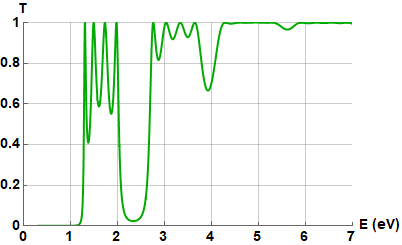}\label{f4.a}
 	}\hspace{1cm}
 	\subfloat[$k_y=0.3\ \AA^{-1}$]{
 		\centering
 		\includegraphics[scale=0.5]{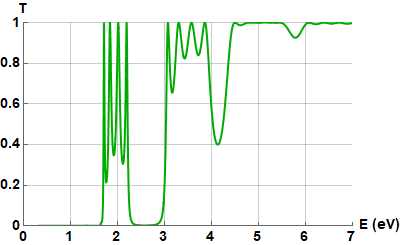}\label{f4.b}
 	}\\
 	\subfloat[$k_y=0.8\ \AA^{-1}$]{
 		\centering
 		\includegraphics[scale=0.5]{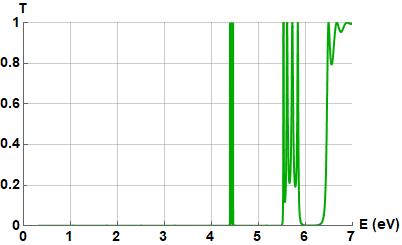}\label{f4.c}
 	}
 	\caption{The transmission  $T$ versus the incident energy $E$ for $V_B=2$ eV, $V_W=0 $ eV, $d_B=d_W= 5\ \AA$ and $n=5$. \textbf{\color{red}{(a)}}: $k_y=0\ \AA^{-1}$, \textbf{\color{red}{(b)}}: $k_y= 0.3\ \AA^{-1}$, \textbf{\color{red}{(b)}}: $k_y=0.8\ \AA^{-1}$. }\label{f4}
 \end{figure}

In Fig. \ref{f4}, we show the transmission $T$ versus the incident energy $E$ for $n=5$ and three distinct values of $k_y$ using the same physical parameters as in Fig. \ref{f3}.  As we can see, the transmission is clearly affected by $k_y$. In fact, at normal incidence, we have some oscillations and the appearance of some pseudo gaps. We note that this behavior is similar to that found for gated graphene superlattices \cite{ref56}. As we demonstrate in Fig. \ref{f4.c}, by raising $k_y$, it is evident that the oscillation zones' widths are reduced while the transmission gaps' corresponding ones are increased. Moreover, once $k_y$ increases, the position of the transmission gaps is shifted toward  the right and the edges become steeper. It is significant to observe that $k_y$ has control over the location and size of the transmission gaps.  Also, the incident angle can be used to control the position and width of the transmission gaps in a similar way to graphene, as a result of its relationship to $k_y$. We note that the application of these transmission gap properties in phosphorene superlattices may aid the creation of numerous phosphorene-based electronics. 

\begin{figure}[ht]
	\centering
	\subfloat[]{
		\centering
		\includegraphics[scale=0.4]{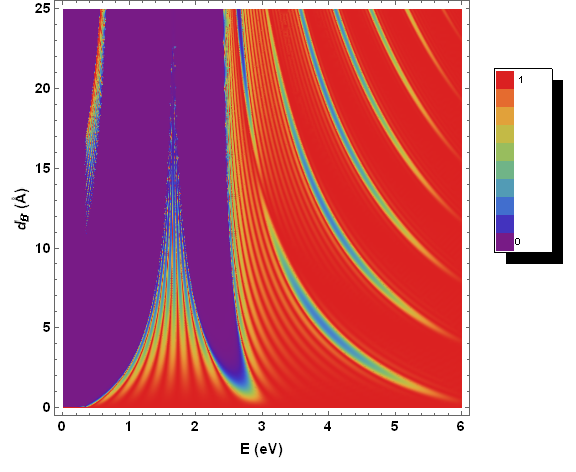}\label{f5.a}
	}\hspace{1cm}
\subfloat[]{
		\centering
		\includegraphics[scale=0.4]{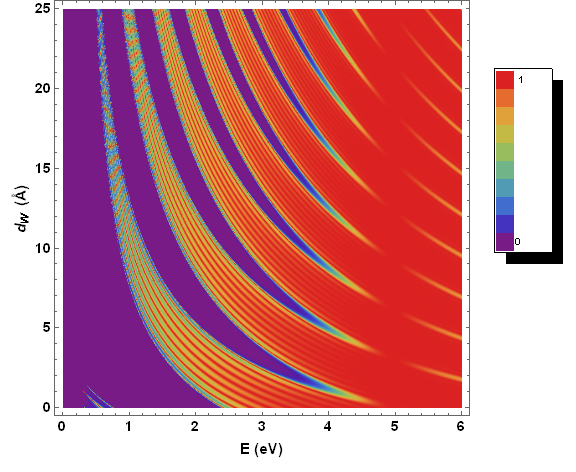}\label{f5.b}
	}
	\caption{Density plot of transmission  $T$ versus the incident energy $E$ and potential width $d_B $ in \textbf{\color{red}{(a)}} with  $d_W=5 \  \AA$  and as a function of $d_W $ in \textbf{\color{red}{(b)}} with  $d_B=5 \  \AA$, for $n=10$, $V_B=2$ \ eV, $V_W=0 \ eV$ and $k_y=0$.
	}\label{f5}
\end{figure}

The transmission constant $T$ is plotted against the incident energy $E$ and potential widths ($d_B$ and $d_W$) in Fig. \ref{f5}. When the well width $d_W$ is constant, there are two remarkable transmission gaps that get closer to each other as the barrier width increases (see Fig. \ref{f5.a}). Moreover, by increasing $d_B$, the width of these two transmission gaps increases. In addition to that, we can clearly see that by increasing the barrier width, some transmission gaps appear for different energies then disappear. It is important to note that these behaviors are similar to those found for monolayer phosphorene in the presence of single \cite{Sarkar17} and bouble barrier \cite{Seffadi22}. In Fig. \ref{f5.b}, the barrier width, $d_B$, is constant. When $d_W=0$, the periodic structure collapses into a single barrier, resulting in a transmission gap, as described in \cite{Sarkar17}. The width of the transmission gap diminishes as the well width $d_W$ increases, and many pseudo-gaps form. Furthermore, the pseudo-gaps become actual transmission gaps when $d_W$ rises.  In addition, as the well width expands and the center of each one shifts to the left, the transmission gaps get nearer to one another. It is significant to note that this implies that by enlarging the well, more narrow channels of an electron wave filter can be generated.  We point out that these outcomes resemble those discovered for graphene superlattices with periodic potential patterns \cite{ref55}.

\begin{figure}[ht]
	\centering
	\subfloat[$n=1$]{
		\centering
		\includegraphics[scale=0.5]{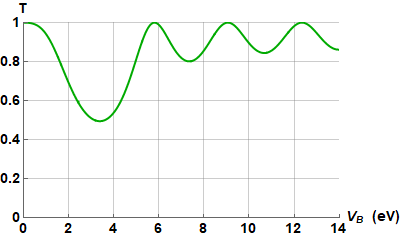}\label{f55.a}
	}\hspace{1cm}
\subfloat[$n=5$]{
		\centering
		\includegraphics[scale=0.5]{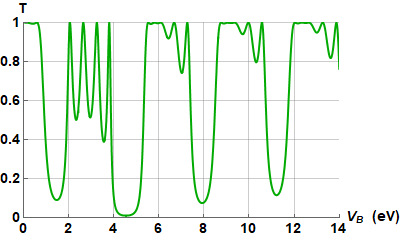}\label{f55.b}
	}\\
\subfloat[$n=30$]{
		\centering
		\includegraphics[scale=0.5]{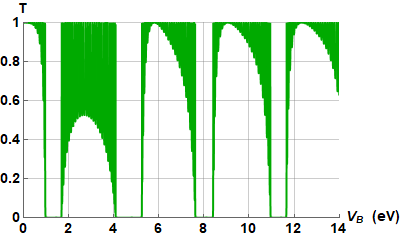}\label{f55.c}
	}
	\caption{The transmission coeficient $T$ as a function of $V_B $, with $k_y=0$, $d_B=d_W=5 \ \AA$, $V_W=0$ and $E=2$\ eV. \textbf{\color{red}{(a)}}: $n=1$, \textbf{\color{red}{(b)}}: $n=5$,
		\textbf{\color{red}{(c)}}: $n=30$.
	}\label{f55}
\end{figure}

Fig. \ref{f55} depicts the transmission function $T$ versus the barrier height $V_B$ for various periodic potential structures with $n=1,5,30$. We can clearly see that by increasing $V_B$, some pseudo-gaps appear with some oscillations around them. As the number of cells increases, the pseudo-gaps become deeper and the edges steeper until they turn into real transmission gaps, and the number of oscillations also increases. As a result, it is possible to control the number of transmission gaps by changing the barrier width. We see that for large $n$, the width of the transmission gaps is unaffected by the quantity of cells and remains constant.

\begin{figure}[ht]
	\centering
	\subfloat[$ k_y=0 \ \AA^{-1}$]{
		\centering
		\includegraphics[scale=0.3]{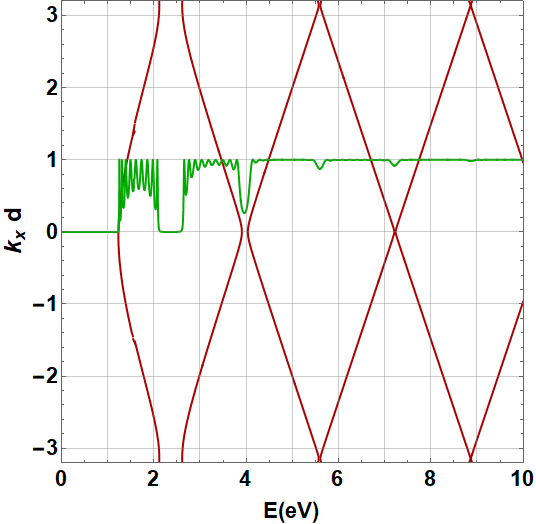}\label{f6.a}
	}\subfloat[$ k_y=0.3 \ \AA^{-1}$]{
		\centering
		\includegraphics[scale=0.3]{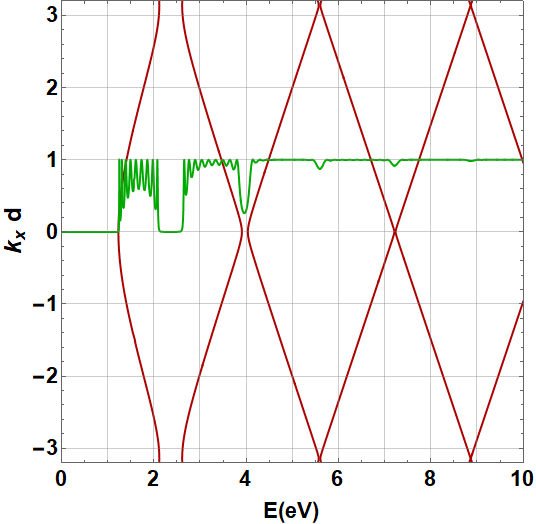}\label{f6.b}
	}
\subfloat[$ k_y=0.8 \ \AA^{-1}$]{
		\centering
		\includegraphics[scale=0.3]{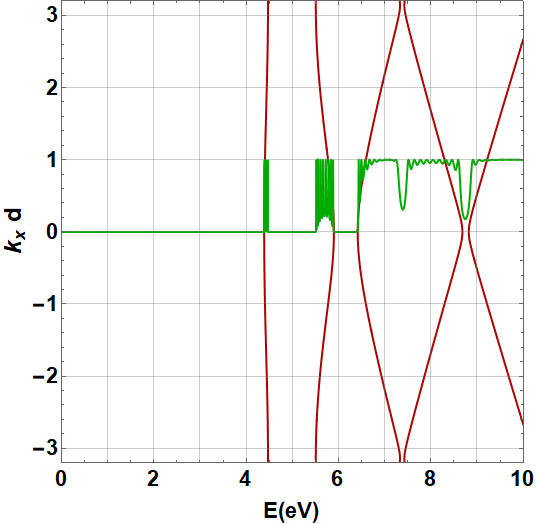}\label{f6.c}
	}
	\caption{The dispersion relation (\text{\ref{f0004}}) versus $k_x d$ (red lines) and the transmission  $T$ versus the incident energy $E$ (green lines) for $V_W=0$, $V_B = 2$ \ eV, $n=10$ and $d_W=d_B=5\ \AA$. \textbf{\color{red}{(a)}}: $ k_y=0$, \textbf{\color{red}{(b)}}: $ k_y=0.3 \ \AA^{-1}$,
		\textbf{\color{red}{(c)}}:  $ k_y=0.8 \ \AA^{-1}$.
	}\label{f6}
\end{figure}

Fig. \text{\ref{f6}} depicts the dispersion relation versus $k_x d$ and the transmission  $T$ versus the incident  energy $E$ for $n = 10$, 
$ V_B=2$ eV, $V_W=0$\ eV, $d_W=d_B=5\ \AA$, and three distinct values $k_y=0, 0.3, 0.8 \ \AA^{-1}$. We can easily observe from Fig. \ref{f6.a} that there is no Klein tunneling at normal incidence since there is a forbidden gap.
This is in contrast to gated graphene superlattices, which do not have a forbidden gap \cite{ref55}. The electronic band also features minibands that are spaced apart from one another by band gaps.  The position and the width of these gaps depend on the transverse wave vector $k_y$. Indeed, when $k_y$ increases, their positions and widths change, moving to the right.  Furthermore, it is seen that the electronic structure's band gaps and transmission gaps match up. 

\begin{figure}[H]
	\centering
	\subfloat[$n=10$]{
		\centering
		\includegraphics[scale=0.4]{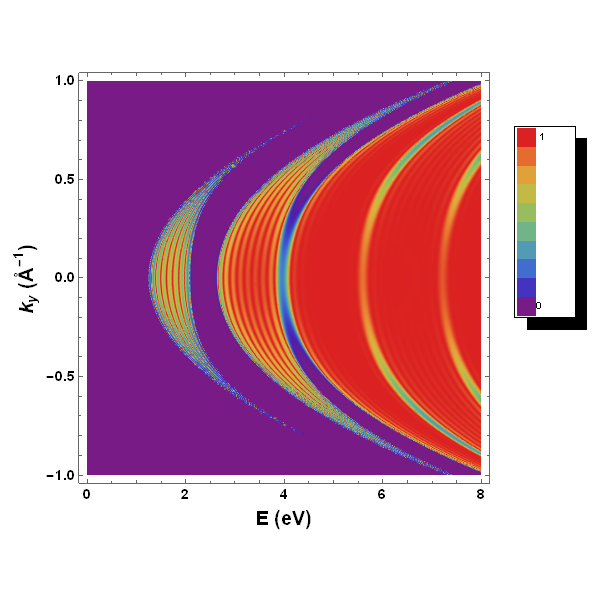}\label{f7.a}
	}\ \ \ \
\subfloat[$n=60$]{
		\centering
		\includegraphics[scale=0.4]{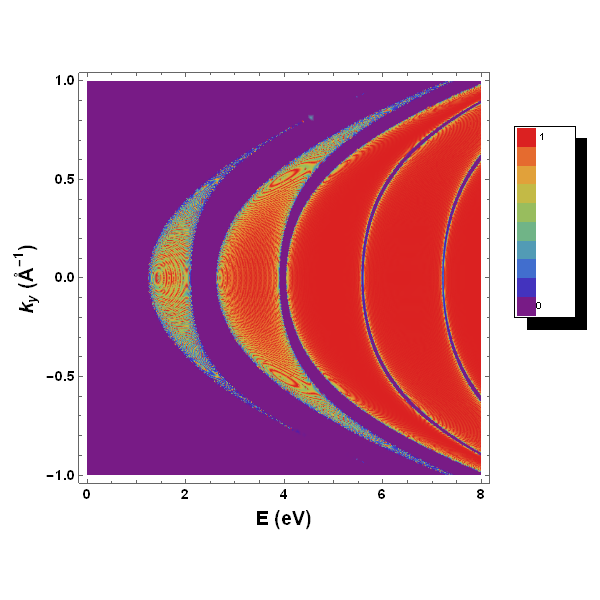}\label{f7.b}
	}
	\caption{Density plot of transmission $T$ verse the transverse wave vector $k_y $ with $V_B=2$\ eV, $V_W=0$ and $d_B=d_W=5 \  \AA$. \textbf{\color{red}{(a)}}: $n=10$, \textbf{\color{red}{(b)}}:  $n=60$.
	}\label{f7}
\end{figure}

For 10 and 60 cells, Fig. \ref{f7} shows a density plot of the transmission $T$ as a function of the transverse wave vector $k_y$ and incident energy $E$.  It is clearly seen that the pseudo-gaps and real transmission gaps are affected by the number of  cells and $k_y$. In fact, by increasing the number of periods, some pseudo-gaps become real transmission gaps. Moreover, by increasing $k_y$ the number and the width of the transmission gaps change. When the superlattice structure is fixed, the effect of $k_y$ can be used to control a desired transmission gap. The number of oscillations beyond the transmission gaps is also influenced by the number of periods. In fact, there are more oscillations on both sides of the transmission gaps as there are more cells.  This can be explained by the phenomenon known as multiple reflection, which is caused by each additional cell and is a result of the Fabry-Perot model \cite{Masir17,Shytov08,Young09}. Furthermore, increasing  $k_y$ widens the transmission gaps and shifts the center of each one toward the right step by step. In addition, as $k_y$ increases, the distance between adjacent transmission gaps decreases. We see that the transmission gaps combine into one transmission gap with a wide width for a large value of $k_y$. Similar outcomes were obtained for gated graphene superlattices \cite{ref55}.


\section{Conclusion }\label{Sec4}

We have looked into the transmission that charge carriers may pass through monolayer phosphorene superlattices. This is made up of periodic potentials involving barriers and wells to generate identical cells. 
As a first step, we wrote down the Hamiltonian model that best describes the present  system and obtained the relevant energy bands.
We have calculated the dispersion relation using Bloch's theorem and
 then looked into the transmission  by employing the transfer matrix approach. These findings revealed intriguing characteristics that were examined under various physical parameter circumstances. 

According to our numerical analysis, there is no transmission gap when $n=1$ (one cell), which is equivalent to one barrier. More than one pseudo-gap is seen for the situation $n>1$.  These pseudo-gaps get deeper and have steeper edges as $n$ rises, eventually developing into genuine transmission gaps.  Additionally, as $n$ rises, more oscillations emerge from the transmission gaps until they are fully transmitted.  Furthermore, the ratio between the number of resonances and the number of cells is constant.  We have also seen that increasing the transverse wave vector $k_y$, barrier height, barrier, and well width increases the number of transmission gaps.  In addition, the width of the transmission gap can be controlled by adjusting $k_y$ and the potential width. The transmission gaps get closer to each other as the well width increases and the center of each one moves left, which means that more narrow channels of electron wave filter can be obtained by increasing $d_W$. We notice that the results are similar to those obtained for gated graphene superlattices. Moreover, we have obtained that the transmission gaps coincide with the band gaps in the electronic structure. However, at normal incidence there is a forbidden gap, which means that there is no Klein tunneling effect, in contrast to the case of graphene superlattices. It should be noted that the findings can be used to develop a variety of phosphorene-based electronic devices, such as electron wave filters.

	
\end{document}